\documentclass[a4paper,11pt]{article}
\usepackage{latexsym}	
\usepackage{amsmath}
\usepackage{amsthm}

\usepackage{ amssymb }
\def\ba{\begin{eqnarray}}
\def\ea{\end{eqnarray}}

\def\ba{\begin{eqnarray}}
\def\ea{\end{eqnarray}}

\def\lb{\label}
\def\be{\begin{equation}}
\def\ee{\end{equation}}

\theoremstyle{plain}

\begin{document}
\baselineskip0.25in

\title{Electric and magnetic axion quark nuggets, their stability and their detection}
 \author{ Osvaldo P. Santill\'an \thanks{Instituto de Matem\'atica Luis Santal\'o (IMAS), UBA CONICET, Buenos Aires, Argentina
firenzecita@hotmail.com and osantil@dm.uba.ar.} and Mat\'ias Semp\'e \thanks{Departamento de F\'isica, UBA CONICET, Buenos Aires, Argentina, sempe(underscore)100@hotmail.com.}}

\date {}
\maketitle

\begin{abstract}

The present work studies the dynamics of  axion quark nuggets introduced in \cite{zhitnitsky} and developed further in the works \cite{zhitnitsky2}-\cite{zhitnitsky13}. The new feature considered here is the possibility that these nuggets become ferromagnetic. This possibility was pointed out in \cite{tatsumi} for ordinary quark nuggets, although ferromagnetism may also take place due some anomaly terms found in \cite{son}-\cite{son2}. The purpose of the present 
letter however, is not to give evidence in favor or against these statements. Instead, it is focused in some direct consequences of this ferromagnetic behavior, if it exists. The first 
is that the nugget magnetic field  induces an electric field  due to the axion wall, which may induce pair production by  Schwinger effect. Depending on the value of the magnetic field, the pair production can be quite large. A critical value 
for such magnetic field at the surface of the nugget is obtained, and it is argued that the value of the magnetic field of \cite{tatsumi} is at the verge of stability and may induce large pair production. The consequences of this  enhanced pair production may be unclear. It may indicate that the the nugget evaporates, but on the other hand it may be just an indication that the intrinsic magnetic field disappears and the nuggets evolves to a non magnetized state such as  in \cite{zhitnitsky}-\cite{zhitnitsky13}.
The interaction
of such magnetic and electric nugget with the  troposphere of the earth is also analyzed. It is suggested that the cross section with the troposphere is enhanced in comparison with a non magnetic nugget but still, it does not violate the dark matter collision bounds. 
Consequently, these nuggets may be detected by impacts on water or by holes in the mountain craters \cite{nature}. However, if the magnetic field does not decay before the actual universe, then this would lead to high energy electron flux due to its interaction with the electron gases of the Milky Way. This suggests that these magnetized quarks may be a considerably part of dark matter, but only if their hypothetical magnetic and electric fields are evaporated.

\end{abstract}

\section{Introduction}

The existence of lumps of quark matter has been postulated long ago  \cite{bodmer}-\cite{witten}. This is not to be confused with nuclear matter, which is composed by a large number of protons and neutrons whose main interaction is due to nuclear forces. Instead, quark matter is approximately a Fermi gas composed of $3N_B$ quarks constituting a color singlet baryon of baryon number $N_B$, and their interaction is much weak, due the fact of the quotient of the cross section to the mass of the object $\sigma_n/M$ is very small.  Some of these states of quark matter are composed only by $u$ and $d$ quarks, and these are known as non strange quark matter. Other form is strange matter, which contains $s$ quarks as well as $u$ and $d$ quarks, in such a way that flavor equilibrium is established by weak interactions $d\to u+e+\overline{\nu}$, $s\to u+e+\overline{\nu}$ and $s+u\to d+u$. Strange matter also contain gluons, and a small component of electrons whose role is insure electric charge neutrality \cite{witten}. 
These strange lumps may be formed during an hypothetical phase transition of the early universe
from a quark gluon plasma state to a hadron phase. There is debate about the nature of this transition, it can be a first order or second order one, a crossover or even an spinodal type of transition. But there are segregation scenarios
which predicts the formation of these lumps, irrespective on the order of the phase transition. An example is given in the reference \cite{atreya}.

Strange form of matter has received special attention, since it is conjectured to have lower energy per baryon number than ordinary nuclei, thus it may constitute a stable state \cite{bjorken1}-\cite{bjorken2}. In other words, the energy per baryon is less
than $930$ MeV, at least at very low temperatures. An heuristic argument for stability comes from the fact that a non strange quark lump posses a Fermi momentum  $p_f\sim$ 300-350 MeV, while the strange quark mass is around $m_s\sim $ 80-130 MeV. As the Fermi momentum is very high in comparison, it may be energetically favored for some non strange quarks to become strange, as this conversion may lower the energy and the Fermi momentum of the system. The estimations given in \cite{bjorken1}-\cite{bjorken2} suggest that the inclusion of strange matter decrease the energy per baryon by 50-70 MeV.  The range of baryon numbers allowed for these objects is $10^2<N_B<10^{57}$ \cite{estable1}. The upper bound is due to the fact that higher baryon numbers collapse into a black hole. The lower bound arises due to shell effects which raise the energy per baryon. 

The bounds $10^2<N_B<10^{57}$ have been further constrained. By use of detailed balance arguments, it is found that primordial nuggets with $N_B<10^{52}$ evaporate. However \cite{madsen2} suggested that nuggets with $N_B>10^{46}$ survive hadron emission, and lower charge nuggets also survive due to reabsortion of hadrons. 
Another constraint comes from nucleosynthesis. The point is that, if quark  nuggets exist and are stable, then iron may not be the ground state of nuclear interactions. This is a delicate aspect when studying nucleosynthesis and was initially considered in \cite{applegate}. For instance, for an inflationary universe with $\Omega=1$, the prediction of the density of the $^3$He and D elements is below the observed value. This problem may be solved if there is some unknown mechanism of photodisintegration of $^4$He into this elements. This requires high energy photons which may be produced by decay of massive neutrinos by non standard interactions. Such channels may be sourced by the presence of quark nuggets
through a weak interaction of high order, as in this case the decay rate will be close to zero. Despite these observations, it was suggested in \cite{schaeffer} that quark nuggets with $10^{16}<N_B<10^{22}$ are consistent with a universe with $\Omega=1$ and with the predicted abundances of the light elements. This motivated a special interest for studying the cosmological consequences of these objects, in particular, in dark matter applications.

The mass density of a quark nugget is $\rho\sim 35. \;10^{16}$kg/cm$^3$, and its mass is approximately $6m=N_B$\;GeV. The surface electron cloud for nuggets with $N_B<10^{15}$ extend
to a distance $r\sim 10^{-8}$ cm from the surface while for very high baryon numbers this distance can be around $400$ fm. Such nuggets therefore possess a Coulomb barrier, with a value close to 10 MeV.  Due to this barrier, only neutrons at low energies may be absorbed by the nugget. There is also a probability of emission of a neutron.  
Further stability issues were considered in \cite{estable1}-\cite{estable10}, and it is believed that stable nuggets
have a mass in the range $10^{-8}$ kg$<m_n<10^{20}$ kg. Thus, very large quark nuggets may masses of planetary order.

Several years ago  it was suggested  that quark nuggets may reach a highly magnetized final state \cite{tatsumi}. This observation is partially motivated from the Bloch works about the possible ferromagnetism of an electron gas \cite{bloch}. This hypothesis was supported in \cite{lam}, who showed that for an electron gas in a fully polarized state the ferromagnetic state is stable. These claims were subsequently tested for a quark liquid in \cite{tatsumi} and 
some evidence supporting a ferromagnetic state was collected for quark nuggets. The presence of an intrinsic magnetic field is  an attractive feature, as it may make the nugget more stable if $B<10^{16}$T \cite{chakrabarty}. A problem is that such highly magnetized magnetic fields may violate the collision requirements for dark matter. However, it was shown in \cite{nature} that  due to the fall of the magnetic field as $R^{-3}$, the resulting cross section is acceptable for representing dark matter. 

The present work is devoted to the effect of such magnetic fields in a different type of quark nuggets, which were introduced in \cite{zhitnitsky}, in which the quarks are trapped in the bulk of an axion domain wall. There are several theoretical reasons for which these axion quark nuggets are interesting. First, the axion particle is an attractive candidate for solving the CP problem in QCD \cite{peccei}-\cite{svz}.  However, the formation of axion domain walls is problematic, since they generate an energy density which largely overcomes the critical density. Nevertheless, there are some axion models for which the axion domain structure does not possess this problem \cite{sikivie}.  These domain walls induce a large value of baryon number due to the fermions living on the wall, which can be shown by bosonization techniques \cite{zhitnitsky5}. Due to the surface tension, these wall tend to contract and segregate baryons and anti-baryons. It is likely that the contraction of these objects stops when the surface tension compensate the pressure difference and the resulting bubble enters into the CS phase \cite{cfl}-\cite{cfl2}. These nuggets may be formed without relying on the order of the phase transition from the quark gluon plasma to the confined phase.

The organization of the present work is as follows. In section 2.1 the modified Maxwell equations for magnetized quark nuggets in presence of an axion domain wall are considered. It is argued that neither the magnetic field found in \cite{tatsumi} nor the axion wall profile are considerably deformed by the interaction between each other. The resulting electric field in the nugget is then estimated. In section 2.2 the pair creation by Schwinger effect in the wall is calculated by use of several methods, and it is argued that the magnetic field \cite{tatsumi} is at the verge of stability. In section 3 the fate of a quark nugget impacting in the earth is analyzed and its possible detection is specified. It is also argued that if these nuggets had a density close to dark matter then they would generate a high energy flux due to their interactions with the Milky Way electrons.
Section 4 contains the interpretation of the results.

\section{Ferromagnetic axion nuggets}

\subsection{The equations describing the configuration}

If the arguments of \cite{tatsumi} about a possible ferromagnetic nature of these nuggets are taken into account, then there should be a magnetic field $B$ due to bulk effects inside the object. For standard nuggets, the reference \cite{tatsumi} presents a magnetic field that scales as
\be\lb{magno}
B= \frac{B_0r_n^3}{r^3}.
\ee
Strictly speaking, a nugget does not have a spherically symmetric radial magnetic field. In fact it may depend on the polar $\psi$ angle in spherical coordinates. The value (\ref{magno}) represents a extrema value say, at certain north or south pole drawn on the nugget.
This extreme numerical value at the surface of the nugget $r_n$ is estimated as $B_0\sim 10^{12\pm 1}$T or $B_0\sim 10^{16\pm 1}$Gauss \cite{tatsumi}.  For ordinary quark nuggets, the work \cite{chakrabarty} suggests that such large magnetic field increase the stability of the configuration, if it does not reach a critical value of the order $B\sim 10^{16}$T. 

Consider now a magnetized quark nugget as above, but including an axion domain wall \cite{zhitnitsky}-\cite{zhitnitsky13}. The domain wall profile function is denoted  by $a(r)$, and interpolates
between two axion vacuums $a_0$ and $a_0+2\pi f_aN$ with $f_a$ the axion constant, whose standard values run between $10^{9}$GeV$<f_a<10^{12}$GeV \cite{sikivie}. 
The value of $N$ is usually not far from unity.
The choice of the window of values for $f_a$ has been justified previously as follows.
 Below the QCD temperature $T_{qcd}\sim 100-150$ MeV, there appears an induced periodic
potential $U(a)$, and the axion becomes light but massive. A customary assumption is that the axion is
at some value of the order of the top of the potential $U(a)$ at the time where this transition occurs. When the Hubble constant
is of the same order as the axion mass this pseudo scalar falls to the potential minimum and starts
coherent oscillations around it. The initial amplitude, which correspond to a maximum, is $A\sim f_a$
and thus, the energy stored at by these oscillations is of the order $E\sim A^2 m_a^2$. The authors of \cite{preskill}-\cite{preskill2}
analyzed the evolution of these oscillations to the present universe and found that the axion energy
density today would be larger than the critical one $\rho_c\sim 10^{-47}\text{GeV}^4$ unless the bound $f_a<10^{12}$GeV
takes place.  On the other hand, there are phenomenological observations which fix this scale $f_a>10^{9}$GeV \cite{dicus}. This lower
bound is required for suppressing the power radiated in axions by the helium core of a red giant star
to the experimental accuracy level. Current algebra methods \cite{bardeen} estimate the axion mass by the relation
\be\lb{ca}
m_a\sim \frac{m_\pi f_\pi}{f_a}.
\ee
This fixes the axion mass in the window $10^{-6}$eV$<m_a<10^{-3}$eV.
The axion potential $U(a)$ is given in these terms by
\be\lb{potoncial}
U(a)=\frac{m^2_\pi f_\pi^2}{2}\bigg(1-\cos \frac{n a}{f_a}\bigg).
\ee
Here $n$ takes integer values, which depends on the particular axion model in consideration.

It should be emphasized however, that the previous discussions holds for the standard QCD axions \cite{peccei}-\cite{svz}. There exists several alternatives to this picture. 
An example are axion models which try to avoid the isocurvature problems \cite{lyth}-\cite{lyth2}, examples can be found in \cite{lesgourgues}
and references therein. Nevertheless, the present work is focused in the standard QCD axion  \cite{peccei}-\cite{svz}, as the axion nuggets in  \cite{zhitnitsky}-\cite{zhitnitsky13} seems focused in these 
type of Goldstone pseudo-bosons.

The explicit form of the function $a(r)$ describing the axion domain wall is not too relevant in the following discussion. The important point is that $a(r)$ evolves from one vacuum to the other in a radial distance of the order $$R\sim (0.3-0.5)\frac{N_B^{\frac{1}{3}}}{160 MeV}(10^{-3}-10^{-6}),$$
the last factors take into account the axion mass bound $10^{-6}$eV$<m_a<10^{-3}$eV, which is assumed to hold in the present work. The baryon number range that the references \cite{zhitnitsky}-\cite{zhitnitsky13} employ is $10^{23}<N_B<10^{32}$. For the larger baryon number one has
$10^{-6}$cm$<R<10^{-3}$cm, while for the smallest value $10^{-9}$cm$<R<10^{-6}$cm. The mass bound is given by $10^{-6}$kg$<m_n<10^{5}$kg, thus these objects can be quite heavy.
The value $N_B\sim 10^{25}$ is attractive in the context of the physics of the 511 KeV line emission from dwarf galaxies \cite{zhitnitsky122}, but it will be assumed that the whole range is valid.

In presence of an axion field $a(x,t)$, when a magnetic field is turned on, an electric field is induced and viceversa.
This follows from an inspection of the Maxwell equations in presence of an axion field, which are given by
\begin{equation}
    \nabla\cdot E= -g_{a\gamma\gamma} \nabla\cdot (a B)+4\pi \rho_e,
\label{1}
\end{equation}
\begin{equation}
    \qquad \nabla\times E=-\partial_t B,
    \label{2}
\end{equation}
\begin{equation}
    \nabla\cdot B=0,
    \label{3}
\end{equation}
\begin{equation}
    \qquad\nabla\times B=\partial_tE+ g_{a\gamma\gamma}\nabla a \times E-g_{a\gamma\gamma} B \partial_t a +4\pi J,
    \label{4}
\end{equation}
\begin{equation}
    \square a+\frac{\partial U}{\partial a}=-g_{a\gamma\gamma} E\cdot B.
    \label{5}
\end{equation}
The axion dependent terms  in these equations arise due to the interaction ${\cal L}_a=g_{a\gamma\gamma} a E\cdot B$ between the axion and the electromagnetic field.
Here $g_{a\gamma\gamma}=c\alpha/\pi f_a$ with $\alpha=1/137$  the fine structure constant. The constant $c$ is model dependent, but its value is not far from the unity.
It is seen from (\ref{1}) that the quantity $\rho_a=-g_{a\gamma\gamma} \nabla\cdot (a B)$ can be interpreted as a charge source for the electric field, not present in the standard Maxwell theory. Thus, when a magnetic field
passes through an axion wall, an electric field is induced due to this source. In addition, $U(a)$ is the axion potential defined in (\ref{potoncial}).

From the discussion given above it follows that the magnetic field (\ref{magno}) is modified due to the equation (\ref{4}). Furthermore, an electric field $E$ is induced, as is seen from equation (\ref{1}).  It is explicitly given by
\be\lb{elec}
E=-g_{a\gamma\gamma} a B.
\ee
The presence of the electric field may generate electron positron pairs due to Schwinger effect. The stability of the configuration is then compromised if the resulting electric field (\ref{elec}) is very strong.

Clearly,  the axion profile $a(r)$ of references  \cite{zhitnitsky}-\cite{zhitnitsky13} should be modified in presence of the magnetic field $B$, which is also deviated from its original form (\ref{magno}). In fact, the equation (\ref{2}) shows that $\nabla \times E=0$, as the magnetic field $B$ is static. In other words, the electric lines for the configuration are never closed. This may be non true if $\nabla \times (a B)\neq 0$, as seen from (\ref{1}), unless the profile $a$ and the magnetic field $B(r)$ are adapted for this to happen. Thus, the task of  solving the explicit profiles may be a complicated one. For this reason, it will be assumed  below that the axion domain wall function $a(r)$ keep its shape approximately and that the magnetic field $B(r)$ do not deviate considerably from (\ref{magno}). But before considering the decay of the induced electric field by Schwinger effect, it may be convenient to justify these approximations first.
 
  Consider first the approximation that the magnetic field does not deviate considerably from the functional dependence
  (\ref{magno}).  The term $J_a=g_{a\gamma\gamma}\nabla a\times E$ is the axion induced current in (\ref{3}). If one assumes
  that the field $a$ make a sudden change of $a\to a+2\pi f_a$ at the nugget surface, then the axion surface magnetization is proportional to $M\sim 2\pi f_a g_{a\gamma\gamma} E\times n$ with $n$
  the surface unit normal. The induced magnetic field at the surface is of the order $|B_a|\sim |M|$. But $|M|\sim f_a g_{a\gamma\gamma}|E|$ and, by taking into account that $E=-g_{a\gamma\gamma}a B$, it follows
  that $|B_a|\sim |M|\sim f_a^2 g_{a\gamma\gamma}^2|B|\sim \alpha^2 |B|$. This suggest that the magnetic field $B_a$ induced by the axion current is around four orders of magnitude smaller that the internal magnetic field (\ref{magno}), and thus 
  the axion current does not deviate the value (\ref{magno}) considerably.

  Another heuristic argument for this statement about $B_a$ comes from the study of the magnetic field of an infinite wire with radius $r_w\sim m_a^{-1}$ near its border.
 This magnetic field is $B_w\sim \mu_0 J_w/r_c$ with $r_c$ the standard cylindrical radius. If the field value in the border of the wire is near  $B=10^{12\pm 1}$ TeV, then its current should be $J_w\sim 10^{12\pm 1}$ TeV$m_a$. On the other hand, the axion current is $J_a=g_{a\gamma\gamma}\nabla a\times E$, this follows from (\ref{3}). If the domain wall $a$ is not considerably modified, it should have a size of length $L_a\sim m_a^{-1}$ and therefore, the current is of order $|J_a|\sim  g_{a\gamma\gamma}\Delta a m_a E$. But the equation (\ref{1}) shows that $E=- g_{a\gamma\gamma} a B$ and therefore
  $$
  |J_a|\sim g^2_{a\gamma\gamma} \Delta a^2 m_a B\sim \frac{1}{(137)^2} m_a B.
  $$
  Here the fact that $\Delta a\sim 2N f_a \pi$ has been taken into account, with $N$ an integer not far  from unity.
 This current is fourth orders of magnitude larger than $J\sim 10^{12\pm 1}$ TeV$m_a$, which suggest that the induced magnetic field due to the axion wall is of smaller order than the  magnetic field (\ref{magno}) itself. 
   
   The estimations given in the last two paragraphs assume that the profile $a(r)$ is not strongly modified in presence of a magnetic field. To see that this may be the case, one should check that the term $E\cdot B$ is not considerably larger than $\partial_a U$ in (\ref{5}), otherwise the axion domain wall will be considerably deformed by this new term. Since the potential (\ref{potoncial}) can be approximated for small values of $a$ as  $\partial_a U\sim m^2_a a$ and $g_{a\gamma\gamma}E\cdot B\sim g^2_{a\gamma\gamma} a B^2$, with $B\sim 10^{11\pm 1}$T,  one has to check that
  $$
  m_a^2\sim\frac{m_\pi^2 f_\pi^2}{f_a^2}>g_{a\gamma\gamma}^2 10^{22\pm 2} \text{T}^2\sim \frac{1}{(137)^2f_a^2 \pi^2}10^{22\pm 2} \text{T}^2.
  $$
But $m^2_\pi f_\pi^2=10^{-4}$ GeV$^4$ and $10^{22\pm 2}$ T$^2\sim 10^{-10\pm 2}$ GeV$^4$, and thus the last inequality is true. Therefore, the assumption that the magnetic field $B$ do not deviate from its  form (\ref{magno}), and that the electric 
field $E$ does not induce a considerable deformation of the domain wall is, at least, consistent. This is the approximation to be used below.

   The field (\ref{magno}) is induced by a ferromagnetic behavior of a quark liquid at the bulk of the nugget. On the other hand, the axion wall can be a source of a magnetic field as well. 
 This is understood from an analysis of anomaly terms for axions \cite{son}-\cite{son2}, in presence of a non zero chemical potential, which is the case for a nugget in the Color Superconducting phase \cite{cfl}-\cite{cfl2}. There are two anomaly terms, one is the standard one $L_1\sim a F_{\mu
\nu} \widetilde{F}^{\mu\nu}$, which induces the decay $a\to \gamma+\gamma$ by the ABJ anomaly diagram. However, in presence of a non zero chemical potential $\mu$ there is a further WZW type of anomaly term given by \cite{son}-\cite{son2}
$$
L_2=\frac{e^2C}{4\pi}\mu B\cdot \nabla a.
$$
Here $B$ is an external magnetic field, and $C$ a model dependent constant, but not far from unity. From the fact that $\nabla \cdot B=0$ it is seen that this term is a total derivative, thus it does not contribute to the equations of motion. However, the presence of this term is not trivial. To see this, consider a flat axion domain wall. Then this term induces an extra energy contribution proportional to $\mu B A$ with $A$ the area of the wall. This energy is produced by a magnetic moment by unit area located at the wall, given by
$$
M=\frac{e^2C}{4\pi}\mu.
$$
For a spherical wall, the total magnetic moment may be zero, since the gradient contributions of $a$ at opposite points are equal and of opposite sign. This implies that the leading contribution to the external magnetic field is the quadrupole one. However, for a simple estimation, one may remember that for a magnetized sphere at the south and north poles the magnetic field is $|B|=\mu_0 |M|$, with $M$ its uniform permanent magnetization. If this value is employed as a guide for a magnitude order, then it is mandatory to estimate the value of $M$. By a naive application of formula (57) of \cite{son1} one finds that
$$
M\sim 10^{12}  \frac{T\mu}{1.5 GeV}\frac{\Delta}{30MeV}.
$$
Here $\Delta$ is a gap characterizing the formation Cooper color pairs, by assuming that the evolution of the object is such that the CFL phase is achieved \cite{cfl}-\cite{cfl2}. A typical value for the chemical potential is $1$GeV and $\Delta\sim 50$ MeV. This implies that the magnetic field these references predict may be of the order of $B\sim 10^{12}$T, which is pretty close to the value of \cite{tatsumi}. The physics describing these fields is however different.

  In view of the present discussion, a generic magnetic field will be considered below without relying wether it is induced by the axion wall or by the quark liquid inside the nugget. The working assumption is then that an electric field $E=-g_{a\gamma\gamma} a B$ is induced, and the task is to estimate its decay probability  due to electron positron pair production. 
 
\subsection{Schwinger effect and estimation of the critical magnetic field of the configuration}
In order to study the electron positron creation by the induced electric field $E$, one may avoid the complication of the inhomogeneity of the electric field $E$ and simply assume that $E\sim g_{a\gamma\gamma}a_0 B_0$ with $B_0$ the surface magnetic field of the bubble. This is the roughest possible approximation. In this situation, the vacuum persistence probability is given by
$$
P_0=|<0_i|0_f>|^2=\exp(-\int_M dx^4 w_e).
$$
 Here the rate of pair creation $w$  by a constant electric field $E$ is expressed by the well known Schwinger formula
\be\lb{se}
w_e=\frac{2 e^2 E^2}{\pi^2}\sum_{n=1}\frac{1}{n^2}e^{-\frac{m_e^2\pi n}{e E}}.
\ee
This formula implies that the vacuum transition probability can be written as follows
$$
P_0=|<0_i|0_f>|^2=\exp\bigg(-\frac{2V T\alpha E^2}{\pi^2}\sum_{n=1}\frac{1}{n^2 e^{\frac{m_e^2\pi n}{e E}}}\bigg).
$$
The approximate volume occupied by the field is roughly $V_f\sim 4\pi R^3/3$. The critical field electric field is
\be\lb{critical}
E_c\sim \frac{m^2_e}{e}.
\ee
If the critical field is reached, then evaporation in electron positron pairs may be relevant. 
This will be the case when
$$
E=\frac{4cB}{137}=\frac{m_e^2}{e}.
$$
Taking into account that $|E|\sim g_{a\gamma\gamma} f_a|B|$ happens when the numerical value of the magnetic field at the surface is $B\sim 10^{11}$ T. 

The conclusion given above is avoiding some subtle point. From the relation $E=-g_{a\gamma\gamma}a B$ it follows that the magnetic field and the electric field of the bubble are collinear.
In this situation, it is known that the magnetic field $B$ suppresses pair creation for scalar fields if it is large enough, but enhance it for fermions \cite{tanji}-\cite{su}. Thus, the pair production
corresponding to the value $B$ found above may be too large. To see that this is not the case, recall that
the rate of electron positron pair creation 
when the collinear magnetic field is turned on is given by \cite{tanji}
$$
w_m=\frac{2 e^2 E B}{\pi^2}\sum_{n=1}\frac{1}{n}e^{-\frac{m_e^2\pi n}{e E}}\coth \frac{\pi nB}{E}.
$$
From this formula it is seen that a large magnetic field increases $w$. In the present case, the magnetic field is large, but it is related to the electric field by $E=-g_{a\gamma\gamma}a B$
with $a$ taking values close to $2\pi$. By taking into account that $g_{a\gamma\gamma}=c\alpha/\pi f_a$  it follows that
$E\sim \alpha B$, with $\alpha$ the fine structure constant. The rate given above is then
\be\lb{sm}
w_m=\frac{274 e^2 E^2}{\pi^2}\sum_{n=1}\frac{1}{n}e^{-\frac{m_e^2\pi n}{e E}}\coth(137 \pi n).
\ee
The leading term of (\ref{sm}) and (\ref{se}) represent the pair production rate by unit time and unit volume.
These are
\be\lb{comp}
N_e=\frac{2 e^2 E^2}{\pi^2}e^{-\frac{m_e^2\pi }{e E}},\qquad N_m=\frac{274 e^2 E^2}{\pi^2}e^{-\frac{m_e^2\pi}{e E}}\coth(137 \pi).
\ee
Now, a rough approach for estimating  the value of the field $E$ for which pair creation is significant may be to evaluate the first rate (\ref{comp}) at the value $E_c=m_e^2/e$ and find the electric field $E$ such
that the second rate has the same numerical value. 
The cotangent factor in (\ref{sm}) is close to the unity and can be neglected. Then the numerical relation defining $E$ is thus given by 
$$
137 E^2 e^{-\frac{E_c}{E}}= E^2_c e^{-\pi}.
$$
The last equation can be cast in the Lambert form
$$
\frac{E_c}{2E}  e^{\frac{E_c}{2E}}=\frac{1}{2}\sqrt{137 e^\pi}.
$$
From here, the solution is given in terms in the Lambert function $W(x)$ as follows
$$
\frac{E_c}{2E}=W(\frac{1}{2}\sqrt{137 e^\pi})\sim 2.44.
$$
This means that $E\sim E_c/4.5$. Thus, the magnetic field does not change considerably the value of the electrical field for which pair creation is appreciable. For this reason, the presence of the magnetic field when studying pair creation can be neglected in this specific case, even though the nugget magnetization is quite large.

The estimation made above assumes that the electric field is uniform in the volume $V_f\sim 4\pi R^3/3$. But this estimation may be not accurate, as the field
$E$ is varying inside the wall. Thus, it is of interest to study the effect of the inhomogeneity of the field inside the wall.  A possible approach is to assume 
that the radius of the bubble is large enough and to analyze what happens at the pole. There are several works that derive results when such inhomogeneities are present \cite{dunne1}-\cite{kleinert2}, and in the following
these references will be followed closely. In order to study the role of the inhomogeneities, the electric can be approximated by a one dimensional one $E=E(z)$ where the axis $\hat{z}$ connects
the pole to the center of the sphere. The field to be considered is of the Sauter form
 \be\lb{sauter}
 E=E_0\;\text{sech}^2\frac{z}{R}.
 \ee
 This choice is simply for convenience, as Schwinger pair creation is understood for this types of potentials  \cite{dunne1}-\cite{kleinert2}. We have investigated other type of shapes, as shown in the appendix.
 However, the results are mathematically more complicated and we do not believe that they change qualitatively the present result. One of the reason for believing that is that
the Sauter electric field (\ref{sauter}) is localized in a region of width $R$, and this imitates the field living on the axion wall bulk. 
 Of course the functional form postulated above correspond to a flat situation and not to a spherical one. But for the sough estimation it may be enough.
  An important point is that, if the instanton method is to be used for estimating the pair production, the two conditions $eE R>>m_e$ and $m_eR>>1$ should be fulfilled. The second is immediately satisfied since
 $l_e\sim m_e^{-1}\sim 10^{-13}$cm, which is much smaller that the nugget radius estimated above. The first condition can be rewritten as $eE>>1/R l_e$.
 But for electric fields near to the critical value one has $eE\sim 1/l_e^2$ and therefore $eER\sim R/l_e^2>>m_e$, as $l_e<<R$. Thus, both conditions are satisfied for fields with values close or larger than the Schwinger critical field.
In this situation, there exists a formula for the pair production per area \cite{nikishov} which, applied to the present case, gives
  $$
 \overline{N}_f=\frac{2(e E_0)^3}{2\pi^2 m_a^3}\int\int_R \bigg(\frac{\cosh Zy-\cosh Zx}{\cosh Z-\cosh Zx}\bigg)(y^2-x^2)dx dy,\qquad Z=\frac{2\pi e E_0}{m_a^2}.
 $$
 The region $R$ of integration is given by
 $$
 R=\{(x,y)| -1\leq -y\leq x\leq y\leq 1,\;\; \epsilon^2\leq (1-x^2)(1-y^2)\},\qquad\epsilon=\frac{m_e m_a}{eE_0}.
 $$
 In order to understand if the spatial inhomogeneity decreases pair production, assume that  the electric field is such that $\delta=e E_0/\pi m_e^2>>1$. This is equivalent to say that the pair production becomes
 significant when the value of $E_0$ is considerably larger the critical field (\ref{critical}). Then the last formula can be approximated by
 $$
 \overline{N}_f\sim \frac{(eE_0)^{\frac{5}{2}}}{6\pi^3 m}.
 $$
 In these terms, one may calculate a value $E_0$ such that the last expression is equal to the value of $N_e$ in equation (\ref{comp}) evaluated at $E_c$.
 This calculation yields the following result
 $$
 \frac{e^\pi}{3} \bigg(\frac{E_0}{E_c}\bigg)^2\sqrt{\delta}=1.
 $$
But this equation should be applied only for  $\delta>>1$, as follows from the assumption that $E_0>> \pi E_c$. This condition contradicts the last equation. This implies that $\delta\sim 1$  and therefore the sough field can not be much larger than $E_c$. This reasoning suggest that the spatial inhomogeneity does not alter considerably mean value of pair production on the axion wall.
 
 The result given are somehow similar to  one related to scalar pair production for Sauter fields and found in \cite{dunne1}, but generalized to fermions. In that reference, it is found that the quotient of the rate of pair creation for a Sauter field $w_s$ and for a uniform electric field $w_e$ is given by
 $$
 \frac{w_s}{w_e}=(1-\gamma^2)^{\frac{5}{4}} e^{-\frac{m^2_e}{\pi e E}\bigg[\frac{2}{1+\sqrt{1-\gamma^2}}-1\bigg]},\qquad \gamma=\frac{m_e}{eER}.
 $$
 This quotient shows that the rate is suppressed due to the inhomogeneity parameter $\gamma$, which tends to zero when $R\to \infty$. But the inhomogeneity parameter for the quark nugget with size $R$ considered here is much less than unity, and there is no significant numerical difference between $w_s$ and $w_e$. The results of the previous paragraph generalize partially this result to the case in which there is electron positron pair production instead of scalar fields.

The above discussion suggest hat a magnetic field of $B\sim 10^{11}$T may induce pair production considerably. The value of the reference \cite{tatsumi} is $B\sim 10^{12\pm 1}$T, which shows that these magnetic field is at the very verge for stability. The magnetic field of the references \cite{son}-\cite{son2} are also of this order. Thus, there may be considerably pair production, unless the combined physics axion quark gluon plasma does considerably change the value of the magnetic field. 

It is not clear for us what is the real fate of such nugget. These apparent instability may indicate that the nugget evaporates, or simply that the ferromagnetic state is not effective due to pair decay of the electric, and thus the magnetic field disappears. In the following, the second possibility will be considered. In this case, the nugget will evolve to the state described in \cite{zhitnitsky}-\cite{zhitnitsky13}

\subsection{Further comments about the approximations made}
 There are some approximations made during the previous analysis, such as (\ref{elec}), which were justified by use of the classical Maxwell equations (\ref{1})-(\ref{5}).
 However, due to the strong nature of the electromagnetic field considered, there are quantum corrections to be added to Maxwell equations. 
 This is due to the fact that the influence on the quantum vacuum of these fields may generate non linearities, which should be properly included in the scenario.
 These linearities are an active research area, and have several interesting effects. First, the photon polarization tensor in presence of homogeneous electromagnetic external fields was studied in \cite{karbstein1}
 and \cite{karbstein33}. The effect of slowly varying inhomogeneities was also considered for instance in \cite{karbstein7}, and the magnetic response in presence of Coloumb sources was studied in \cite{karbstein11}. In addition, the presence of magnetic fields produce an enhancement at the light shinning through wall experiments designed in  \cite{karbstein2}-\cite{karbstein3}, which may be seen in a near future. Further applications of these non linearities for laser fields were considered in \cite{karbstein6}. In addition, the interaction between micro-bubbles with ultra intense laser pulses, by taking into account the QED vacuum polarization, was studied in \cite{karbstein10}. A research of QED vacuum polarization in presence of axion fields was initiated in \cite{karbstein8}.  These result may have interesting cosmological applications.

 The fact that the physics of QED non linearities is so rich raises an issue about the affirmation made above that the nugget electric field $E_n>E_c$, as it may be the case that the non linearities change its numerical value and render it subcritical $E_n<E_c$. We will argue below that this is not the case. In order to see these effects, recall that the exact one loop QED effective lagrangian is given by
 \be\lb{elag}
 {\cal L}_{HE}={\cal L}_{m}+{\cal L}_{a\gamma}+\frac{1}{2(2\pi)^2}\int_0^\infty \frac{ds}{s^3}\bigg[e^2 \epsilon \beta s^2 \coth(e\epsilon s)\cot(e\beta s)-1-\frac{e^2}{3}(\epsilon^2-\beta^2)s^2\bigg]e^{-is(m_e^2-i\eta)}.
 \ee
 This is the well known Euler-Heisenberg lagrangian.  Here ${\cal L}_m$ is the Maxwell lagrangian, ${\cal L}_{a\gamma}$ is the axion photon coupling defined above and the following quantities have been introduced
  $$
  \epsilon=\sqrt{\sqrt{S^2+P^2}+S},\qquad \beta=\sqrt{\sqrt{S^2+P^2}-S},
  $$
  $$
  S=-\frac{1}{4}F_{\mu\nu}F^{\mu\nu}=\frac{1}{2}(|E|^2-|B|^2),\qquad P=-\frac{1}{4}F_{\mu\nu}\widetilde{F}^{\mu\nu}=E\cdot B.
  $$
As is well known, the Euler-Heisenberg lagrangian contains a real part and an imaginary one. The imaginary part describes pair creation, which was analyzed in previous sections. However, the real part describe corrections to the Maxwell equations due to quantum effects. 

It is difficult to evaluate the integral (\ref{elag}) explicitly. However, in some limits, its approximate form is known. For instance, the fields satisfying the condition $\epsilon<<E_c$ and $\beta<<E_c$ are known as weak fields. In this limit, the real part of the lagrangian (\ref{elag}) may be approximated by (see \cite{expansion1}-\cite{expansion3}
and references therein)
  \be\lb{elag2}
 {\cal L}^R_{HE}\simeq{\cal L}_{m}+{\cal L}_{a\gamma}+\frac{2\alpha^2}{45m_e^4}(4S^2+7P^2)+\frac{64\pi \alpha^3}{315 m_e^8}(16S^3+26S P^2).
 \ee
 The modified Maxwell equations that follow from (\ref{elag2}) are
 $$
 \nabla\cdot B=0,\qquad \nabla \times E=-\frac{\partial B}{\partial t},
 $$
 $$
 \nabla \cdot \bigg[\bigg(\frac{1}{2}+\frac{8\alpha^2}{45m_e^4}S+\frac{64\pi \alpha^3}{315 m_e^8}(24S^2+13P^2)\bigg)E\bigg]=4\pi \rho_e
 -\nabla \cdot \bigg[\bigg(g_{a\gamma\gamma}a+\frac{28\alpha^2}{45m_e^4}P+\frac{64\pi 52\alpha^3}{315 m_e^8}SP\bigg)B\bigg],
 $$
 $$
 \frac{\partial}{\partial t}\bigg[\bigg(\frac{1}{2}+\frac{8\alpha^2}{45m_e^4}S+\frac{64\pi \alpha^3}{315 m_e^8}(24S^2+13P^2)\bigg)E+\bigg(g_{a\gamma\gamma}a+\frac{28\alpha^2}{45m_e^4}P+\frac{64\pi 52\alpha^3}{315 m_e^8}SP\bigg)B\bigg]
 $$
 $$
 =-\nabla \times \bigg[\bigg(\frac{1}{2}+\frac{8\alpha^2}{45m_e^4}S+\frac{64\pi \alpha^3}{315 m_e^8}(24S^2+13P^2)\bigg)B+\bigg(g_{a\gamma\gamma}a+\frac{28\alpha^2}{45m_e^4}P+\frac{64\pi 52\alpha^3}{315 m_e^8}SP\bigg)E\bigg].
 $$
 In order to see if these corrections suppress the value of the electric field $E_n$,
 we have checked numerically that given a constant solution $B\sim 10^{11}T$, the resulting electric field is $E>E_c$, so it seems that quantum corrections do not suppress it considerably. The fields where assumed to be 
 time independent for simplicity.  In fact, the approximation (\ref{elag2}) assumes that  the weak limit $\epsilon<<E_c$ and $\beta<<E_c$, but it can be seen by inspection of the quantities defined below (\ref{elag})
 that this is not the case for such enormous magnetic field. This means that the approximation (\ref{elag2}) breaks down for $B\sim 10^{11}T$.
 
 The right situation to consider is instead $B>>B_c$ and $E<<E_c$.  Some limit cases are well known. The approximate form of the Euler-Heisenberg lagrangian for large electric magnetic fields $B>>B_c=E_c/e$  and no electric fields is the well known Weisskopf lagrangian \cite{expansion1}-\cite{expansion3}
  \be\lb{elag3}
 {\cal L}^R_{HE}\simeq{\cal L}_{m}+{\cal L}_{a\gamma}+\frac{e^2 B^2}{24\pi^2}\bigg[\log\frac{\pi E_c}{B}+\gamma\bigg]+\frac{e^2B^2}{4\pi^4}\zeta'(2).
 \ee
 Here $\gamma\sim 0.577216$ is the Euler-Mascheroni constant and $\zeta(z)$ is the Riemann zeta function. For large electric fields $E>>E_c$ and no magnetic field the result is \cite{expansion1}-\cite{expansion3}
  \be\lb{elag4}
 {\cal L}^R_{HE}\simeq{\cal L}_{m}+{\cal L}_{a\gamma}+\frac{e^2 E^2}{24\pi^2}\bigg[\log\frac{\pi E_c}{E}+\gamma\bigg]-\frac{e^2E^2}{4\pi^4}\zeta'(2).
 \ee
For the case $E<<E_c$ and $B>>B_c$, which is the situation of interest here, the real part of the lagrangian is described by the following series expansion \cite{expansion1}-\cite{expansion3}
\be\lb{serie}
 {\cal L}^R_{HE}=\frac{1}{2(2\pi)^2}\sum_{n,m=-\infty}^\infty \frac{1}{\tau_n^2+\tau_m^2}[(1-\delta_{m0})J(i\tau_m m_e^2)-(1-\delta_{n0})J(\tau_n m_e^2)].
\ee
Here 
$$
\tau_n=\frac{n\pi}{e\epsilon},\qquad \tau_m=\frac{m\pi}{e\beta},
$$
and the following function
$$
J(z)=-\frac{1}{2}[e^{-z}E(z)+e^z E(-z)],
$$ 
has been introduced, with $E(z)$ is the exponential integral function. Now, if it were the case that $B>>B_c$ and $E<<E_c$, then $2S\sim -|B|^2$ and $P<<-S$.
Thus 
$$
  \epsilon\sim \sqrt{\frac{1}{2}|S|}\frac{|P|}{|S|}\sim E<<E_c ,\qquad \beta=\sqrt{2|S|}\sim B>>B_c.
 $$
 The denominator in (\ref{serie}) grows with $n$ and $m$. The functions $J(z)$ in the numerator tend to zero for such large values. Based on this, one may consider the terms with $n,m=0,1$
 as the leading ones. This leads to the following approximated lagrangian\footnote{In fact, one may use some summations formulas given in \cite{expansion1}-\cite{expansion3} to include much more terms, but for the estimation to be done here we will assume that the first terms are enough.}
 $$
 {\cal L}^R_{HE}=-\frac{1}{2(2\pi)^2}\frac{e^2\epsilon^2}{\pi^2}J\bigg(\frac{\pi^2m_e^2}{e^2\epsilon^2}\bigg)+\frac{1}{2(2\pi)^2} \frac{e^2\beta^2}{\pi^2}J\bigg(\frac{i \pi^2m_e^2}{e^2\beta^2}\bigg)
+\frac{1}{2(2\pi)^2} \frac{1}{\frac{\pi^2}{e^2\epsilon^2}+\frac{\pi^2}{e^2\beta^2}}\bigg[J\bigg(\frac{i \pi^2m_e^2}{e^2\beta^2}\bigg)-J\bigg(\frac{\pi^2m_e^2}{e^2\epsilon^2}\bigg)\bigg].
$$
Still, if $\epsilon<<\beta$, this may be approximated by
$$
 {\cal L}^R_{HE}=-\frac{1}{(2\pi)^2}\frac{e^2\epsilon^2}{\pi^2}J\bigg(\frac{\pi^2m_e^2}{e^2\epsilon^2}\bigg)+\frac{1}{2(2\pi)^2} \frac{e^2\beta^2}{\pi^2}J\bigg(\frac{i \pi^2m_e^2}{e^2\beta^2}\bigg).
$$
By taking into account that $\epsilon \sim E$ and $\beta\sim B$, and that
$$
J(z)\sim -\frac{1}{z^2},\qquad z<<1, \qquad J(z)\sim -\log(z)\qquad z>>1,
$$
it is found that
$$
 {\cal L}^R_{HE}\sim \frac{1}{4\pi^4}\frac{e^6E^6}{ m_e^4}+\frac{e^2B^2}{8\pi^4} \log\bigg(\frac{ \pi^2m_e^2}{e^2B^2}\bigg).
$$
This lagrangian should be added to the standard Maxwell one, together with the photon axion coupling. In similar fashion as above, we have played with the resulting 
equations of motion. We find that, if it is assumed that $E<<E_c$, the corrections to the Maxwell equations are small. On the other hand, if the corrections are small, this implies
that the classical picture is still valid and $E>>E_c$. This contradiction suggest that the assumption of small electric field is not quite right. 

The arguments given above are not complete proofs, and they do not have into account higher loop calculations \cite{karbstein5}-\cite{karbstein4}.
But we conjecture that the effect of the non linearities will not correct the electric field from a classical value $E>>E_c$
to a corrected value $E<<E_c$. 
It is worthy to mention that there exists a discussion in the literature about the possible role of the non linearities in QED for rotationally powered pulsars, with large magnetic fieldsr, as shown in \cite{karbstein9}
and references therein. The reference \cite{karbstein9} argues that these effect are negligible, at least for certain types of neutron stars. Based on these heuristic argumetns, we assume that the non linearities of QED do not change the supercritical value of $E_n$ and that the relation (\ref{elec}) and the estimated values obtained in previous section still apply for axion quark nuggets.
 
 \section{Imprints of a electrified nugget}
 
 \subsection{Enhanced cross section at the troposphere}
 The study of the cross section of a quark nugget passing through the troposphere is important for understanding which methods are suitable for their detection.
 
 It should be recalled that for a non magnetized quark nugget \cite{rujula} it is usually assumed
 that its cross section $\sigma_n$ is given by the cross sectional area of its core mass density. In this terms the following formula for the energy loss of a quark nugget
$$
\frac{dE}{ds}=-\sigma_n \rho v^2, \qquad \sigma_n=\pi\bigg(\frac{3m}{4\pi \rho_n}\bigg)^{\frac{2}{3}},
$$
is found. One may consider velocities of the order $v\sim 250$km/s, which are characteristics of the Sun galaxy rotation. Here $\rho$ is the medium density the nugget is passing through.
The formula given above is inspired by the physics of meteorites. 
The quark nugget density is approximately $\rho_n\sim 10^{18}$kg/m$^3$. This formula allows to analyze if the nuclearity passing through
the earth will accumulate at the crust or if it will pass through the earth. Besides, a nuclearity that pass through the air or water deposit part of its energy in terms of visible light. 
The authors of \cite{rujula} calculate the fraction of the dissipated energy in terms of visible light, by assuming that the light is emitted as a black body radiation from an expanding cylindrical shock wave.
In this terms, they derive a formula for the luminosity of these objects and discuss possible detection mechanisms based on these formulas.

This picture changes when the nugget entering into the earth has a magnetic field $B$ turned on. A first attempt for studying their energy deposition is to consider some known formulas for magnetic objects such as monopoles, which in principle may be also applied to a magnetized nugget \cite{burdin}. However, it may be convenient to generalize the results obtained in \cite{nature} since they take into account that the nugget is surrounded by a plasma when it enters in the troposphere. In this approach, the physics of a magnetized nugget is as follows. When the nugget enters into the 20 km region covered by the troposphere, it finds a neutral medium. There is oxygen and nitrogen in this region. The oxygen binding energy is around $8$ KeV. The Zeeman term that arises from the interaction of these atoms and the nugget magnetic field is given by $L_z\sim \mu_b B$ with $\mu_b\sim 93. 10^{-25}$J T$^{-1}$ the Bohr magneton. It is seen by passing to natural units that $L_z\sim 6.10^{-5}$eV T$^{-1} B$. Therefore the Zeeman energy will have a numerical value close to the keV scale when $B\sim 10^{7\pm 1}$T. By taking into account that  the magnetic field depends on the radius by a law of the form $B\sim r^{-3}$ in the model introduced in \cite{tatsumi} and by taking into account that at the surface $B\sim 10^{11\pm 1}$ T, one may assume that, inside a radius $r\sim (10-10^{2})r_n$, there is enough Zeeman energy for ionizing oxygen and hydrogen. Thus, in this region, the surrounding medium is approximately an ionized plasma. Assume that the nugget has a velocity $v$, which is not relativistic. In the system in which the bubble is at rest there is an incoming plasma with a speed $-v$. The particles tend to make a round trajectory, of Larmor type, when approaching the region where the magnetic field is strong, the plasma pressure does not allows this to fully happen. Thus, there is a surface formed in which the plasma pressure equals the magnetic pressure. The plasma ram pressure is $P_r\sim \rho v^2$ up to a model dependent constant which is not far from unity. The magnetic pressure is $P_m\sim B^2/\mu_0$. By taking into account the dependence $B=B_0 r_n^3/r^3$ is considered, then the condition $P_m=P_r$ shows that the plasma particles are allowed to approach the nugget up to a radius
\be\lb{redio}
r_e\sim \bigg(\frac{B_0^2r_0^2}{\mu_0 \rho_p v^2}\bigg)^{\frac{1}{6}}.
\ee
The cross section then can be approximated in natural units by  \cite{nature}
\be\lb{cs}
 Q_n\sim \pi r_e^2\sim \bigg(\frac{B_0^2r_0^2}{ \rho_p v^2}\bigg)^{\frac{1}{3}}.
\ee
In other words, this cross section is simply the two dimensional area of a disc of radius $r_e$, which is enhanced due to the magnetic field.

At this point, the calculations given above are the ones already presented in \cite{nature}. These calculations assume that the nugget electric field is zero. However, as discussed in previous sections, in presence of an axion domain wall surrounding the object plus a magnetic field, an induced electric field $E_n$ is induced. Thus, one needs to evaluate the effect of the electric field in those formulas. Our estimation is that the formula (\ref{cs}) is not 
significantly changed, and the reasoning goes as follows. The plasma may be considered a fairly good conductor, and it will try to expel somehow this electric field by forming a charged surface, a shield, which distorts the value of $r_e$. There are two electric field contributions, the nugget original field $E_n$ and one the induced plasma $E_p$.
This electric field repels charges of one sign and attracts the others. On the other hand there is a tendency of the plasma to screen the electric field, as the plasma is a good conductor. It is not clear if the presence of the electric field
deforms considerably the  magnetic surface $r=r_e$ found above or not. However, some estimation can be made. Consider the new surface $r=r'_e$, the border to which the charges are allowed to approach the electrified and magnetized nugget. If this surface is though as the boundary of a conductor, the force on a surface element with area $\Delta S$ is $F=\sigma E_c \Delta S$ where $E_c$ is the surface electric field. The electric field of an ideal conductor is normal to the surface and $E_c\sim \sigma/\epsilon_0$ with $\sigma$ the surface charge density. Thus, the resulting  force can be expressed as $F=\epsilon_0 E^2_c \Delta S$, from where it follows the pressure $P=\epsilon_0E^2$.
By extrapolating this picture to the present situation, one may consider a pressure given by $P=\epsilon_0 E^2$ with $E=E_n+E_p$, with both the induced plasma electric field $E_p$.
It may be fairly reasonable to assume that both fields are of the same order on the sought
surface $r=r'_e$. As shown in previous subsections, the electric nugget field is given by
$$
E_n^2=g_{a\gamma\gamma}^2 a^2 B^2.
$$
By taking into account that $g_{a\gamma\gamma}=c\alpha/f_p \pi$ and that $a=f_a \theta$ with $\theta$ an angular variable, it follows that
$$
E_n^2=\frac{c^2\theta^2B^2}{\pi^2 (137)^2}.
$$
The constant $c$ is not far from unity and the angular variable $\theta$ can take values between $0\leq \theta<2\pi N$ with $N$ an integer, it is seen that the electric pressure $P_e\sim \epsilon_0 E^2$ is less or of the
same order of magnitude than the magnetic one $P_m\sim B^2$ for  controlled values of $N$. Thus, it is likely that the region determined by (\ref{redio}) is not significantly deformed, but the charge distribution is modified in order
to screen the nugget electric field. Therefore, it may be considered that the formula (\ref{cs}) applies even for this type of electrified and magnetized nuggets.

It should be remarked that the formula (\ref{cs}) is obtained purely by a classical considerations. It is true if the quantum corrections described in previous section do not alter the order of magnitude
of $r_e$. In the following, we assume that this is the case.

There are observables consequences of the cross section (\ref{cs}), which follows from reference \cite{nature}.  This can be briefly described as follows.
The formula (\ref{cs}) allows to calculate the energy loss of a nugget of mass $M$ and a given velocity of  the order of $250$km/s through the atmosphere and three meters
of material with density $\rho\sim 1120$kg/m$^3$. This simulation is presented in \cite{nature} and it is found that the mass range for nuggets is such that they may to penetrate the atmosphere and deposit $10^3-10^6$J/length. 
For instance, this curve shows that  nuggets with masses between $0.1-1$kg deposit $00.3-50$MJ/m.
This energy deposition is enough to form a shock wave in water which decays fast into an acoustic pulse. The pressure of this pulse can be monitored by three or more time synchronized sensors, which may determine the impact point by interpolation. From this information, one may deduce the energy deposition and, by use of the curves found in \cite{nature}, obtain the mass of the nugget. In order to decrease the risk of obtaining a
false positive, it is desirable to use three sensors in coincidence. 

The measurements requires to understand the energy deposition on water to pressure pulses. For simulations of this, one may use the CTH shock physics code \cite{cth} and the SESAME equation of state in water \cite{sesame}.
Some of these simulation are presented in \cite{nature}. A good location for detection is the Great Salt Lake in USA, since its salinity does not allow animal life which may contaminate acoustic pulses. There are further
technical problems to be solved. The incident flux as a function of the mass is not known, as the quark nuggets are assumed to have a continuum mass distribution. Under the assumption that the events by mass unity
varies inversely with the mass, the authors of \cite{nature} used a 250 hour sensor which collected events at a rate 300/year, thus obtaining a graphic of the quantity of events by mass unity. Their results suggest 
that detectability is possible.
In addition, the conversion of the deposited energy into acoustic energy and the attenuation of the of the acoustic pressure
at the detectors as a function of the distance point should also be characterized, for the Great Salt Lake enviromment.  The authors of \cite{nature} have studied the effects of the minerals of the lake
in the attenuation of higher frequencies of the acoustic pulse and have found that the conversion into acoustic energy has some similarities with TNT explosions. These preliminary experiments
were performed by C55 hydrophones, and further details can be found in the original reference.

The result given above is interesting, since it opens the possibility that these axion nuggets may be found in near future experiments. However, as discussed below, these electrified and magnetized axion  quark nuggets may generate some high energy rays that are not observed due to their interaction with electron gases in the Milky Way. In this sense, the fact that the electric field is larger than the critical one $E_c$ may give an explanation of this absence, since this field may have decayed before the actual universe.

\subsection{High energy ray generation}

If there were electrified nuggets composing a large part of dark matter, then these nuggets
would interact with electron gases in the Milky Way galaxy. The density of these electron gases
is $n_e\sim\; 1\;\text{cm}^3$. These electrons, after colliding with the nuggets, will accelerate and acquire a energy
$eEm_a^{-1}$. A rough estimation about the number of collision per unit time that a nugget experiences is
$$
n_c\sim R^2 v_e n_e.
$$
Due to the oscillations of the axion field, the electrons acquire an energy $E_e\sim e E/m_a$. These acceleration mechanisms
were discussed in other context for instance in \cite{iwazaki}. This implies that when a astronomical
observation of  $1$ kpc$^3$ is done, the number of nuggets would be 
$$N_n\sim \frac{\rho_d}{m_n}\text{kpc}^3,$$ 
where $m_n$ denotes the mass of the nugget and $\rho_d$ denotes the local density
of the dark matter $\rho_d \simeq 0.3GeV$cm$^{-3}$. The electrons are scattered by
the collisions with the nuggets. The number N of the electrons
scattered per unit time in the area is given by $$N=\frac{R^2v_e n_e
\rho_d}{m_n}\text{kpc}^3.$$
These electrons possess the very high energy
$eE/m_a$ when the electric field is the strong one discussed in the
present paper. For instance, when R is $10^{-9}$cm and $m_a=10^{-3}$eV,
then, the electron energy $eE/m_a\sim 10^4$GeV and the number of the
electrons N is  $10^{20}$/s for baryon number $N_B=10^{28}$. This is for a velocity $v_e\sim 100$km/s. If such a huge number of the electrons with
high energy $10^4$GeV were emitted in the area with the volume $1$ kpc$^3$,
they would be already detected. Thus, it is likely that these nuggets may exist today only in a non magnetized state or, if they are magnetic and electric, 
their density should be considerably small than dark matter.

Therefore, if the nuggets reach a non electrified neither magnetized state, the detection of water impacts discussed in the previous
subsection become more involved, as the cross section becomes considerably smaller. However, there are for now several searches of standard quark nuggets and it may be interesting to see if they apply to the present case.  
An interesting  possibility is the detection at the Alpha Magnetic Spectrometer \cite{alphamagnetic1}-\cite{alphamagnetic4}, which is sensitive to a wide range of mass and charge for these objects. In fact, if these objects reach the non magnetized neither electrified state, they may be similar to strangelets, which have been intensively searched at this spectrometer. However, there is an important detail to be remarked. One assumption for detection for strangelets is that neutron stars are strange stars. If this is so, when two neutron stars form a binare system  which loss energy due to gravitational radiation, they finally collide, and this results in a strangelet cosmic ray. The galactic coalescence of our galaxy is approximately one even every 10.000 years. Simulations estimate that the release of $10^{-10}$ solar masses per year. On the other hand, it is not clear if a considerable axion wall quark nugget  flux may be obtained from such binare collisions, as it is not clear if such axion walls would develop.  If the axion quark nuggets are assumed to be the main component of dark matter, then the flux is given by \cite{zhitnitsky3}
$$
\frac{dN}{dA dt}=nv\sim \frac{10^{25}}{B}\frac{1}{\text{km}^2 \text{yr}}.
$$
This flux is below the sensitivity of the standard searches for dark matter, but is close to the flux of cosmic rays above the GKZ bound. Thus, a further possibility is to consider air showers \cite{showers} that these objects may generate. The shower that is generated by these objects have some similarities with the ones
produced by a single primary with ultra high energy. But it is precisely their subtle differences which may play a significant role in detection. When an anti-nugget hit the atmosphere, the dominant interaction
is strong mediated matter anti-matter annihilation. The resulting shower will be mainly composed by light mesons and their decay products. There are electrons and positrons produced, which are unable to escape
the nugget. However, the muons are able to escape, loose energy to the atmosphere, generate fluorescence light before decay into electron products. This leads of a number of charged particles as a function
of the height, which grown to a maximum particle and then decreasing rapidly, and is very similar to the one produced by a single primary.  However, due to the fact that the nugget is not relativistic in comparison, the nugget shower
fluorescence will elapse more time. The reason is that the matter produced by the nugget is confined to a region few kilometers around the object, and the combined system moves slowly. The time scales involving the lateral surface profile are also larger.  A further characteristic feature is that the lower velocity of the primary particles will lead to a correlation between the arrival direction and the earths motion direction with respect to the galaxy. This arrival direction is related to dark matter distribution and as such, does not have any correlation with the galaxy objects. In addition, the muon spectroscopy may distinguish both showers since the muon spectrum for the nuggets have a cutoff at energies of the order of the GeV while for ultra high energy primaries this is not the case. Further characterization of these showers can be found in \cite{showers}.

The differences between ultra high energy may play a very important role in future detection experiments. These include the AUGER observatory \cite{auger} and the Telescope Array \cite{telescopearray} which may detect these showers by their fluorescence dectectors. The reference \cite{pshirkov} is able to put flux limits that may apply to strangelets, axion wall quark nuggets and other objects.  The macro detector at Gran Sasso may also be sensitive to these objects \cite{sasso}. It is interesting to remark that the results presented in  \cite{wandelt} suggest that the quark nuggets are not excluded when their mass is somewhat
larger than $m> 10^{22}$kg. The study  of the constraints that follows from ancient mica exposure, of ages of the order of $10^{9}$ years, is also an interesting lead \cite{salamon}. Another possibility is that these objects may be detected by impact craters \cite{rafelski}. This craters should be compatible with energetic 
events without the presence of meteorites. The group of authors \cite{nature} is searching for such events in County Donegal in Ireland, and have analyzed several reported events in India or Nicaragua, but where unable to find unambiguous results. However, further experiments may be performed and the crater may be distinguished from usual meteorites by the absence of CUDO and meteorite materials.

\section{Discussion}

In the present work a ferromagnetic axionic quark nugget was considered. It was shown that the internal magnetic field induces an electric field on the axionic wall which decays into electron positron pairs if the magnetic field
is above the critical value $B_c=10^{11}$ T, which close but below the range given in \cite{tatsumi}. The possible role of the inhomogeneities of the electric field was also taken into account, but it was shown that the deviation from the standard Schwinger  formula is small and the critical value remains also unaltered.  The fact that the critical value is below the value of \cite{tatsumi} may be in fact healthy. It may indicate not that the nugget is unstable, but that in fact the ferromagnetic state does not take place. The calculation done contains several approximations, and to study the real evolution of the nugget is for sure more complicated. But we conjecture that the ferromagnetic state is not achieved when an axion wall is present. Consequently, the interaction of these nuggets with the electron gas in the Milky Way does not give a huge number of electrons with high energy contradicting the observations. In other words, we suggest that ordinary nuggets may reach a ferromagnetic state, but the ones with axionic walls evolve to a non ferromagnetic state of the form \cite{zhitnitsky}-\cite{zhitnitsky13}, due to the influence of the axion wall during its the evolution.

On the other hand, the energy deposition on the earth due to these objects was studied. For the electrified and magnetized state, this was investigated by generalizing the magnetopause model considered in \cite{nature}. This model follows an analogy between the plasma surrounding the nugget and the solar wind interacting with the magnetic field of the earth. In the present work, the effect of the electric field was included and it was shown, by use of the modified Maxwell equations in presence of an axion, that the electric field pressure on the plasma is less than the magnetic one. Consequently, the cross section for a nugget interacting with a plasma is not considerably modified. A direct application of the results given in \cite{nature} implies that these nuggets, even in the electrified and magnetized state, do not violate the collision bounds for dark matter. However, as we conjectured that the electrified and magnetized state do not take place, alternatives for non electrified neither magnetized nuggets were also briefly discussed. Perhaps the study of air showers is promising, since the products of the decay can be qualitatively different from a single primary with ultra high energy.

\section*{Acknowledgments}
O. S is supported by CONICET, Argentina. We gratefully acknowledge a discussion with A. Ilderton  for providing us some details about his work on pair production, and to F. Karbstein for orientation about effects of non linearities in QED.

\appendix

\section{Electric field from a spherical domain wall with constant magnetic field}
In the present paper, the role of inhomogeneities were estimated in terms of the Sauter field (\ref{sauter}). 
We have worked out some other possible forms of the electric field, but the results are more complicated and do not change much
the presented picture. We collect however in the present appendix some of the formulas that we have obtained
which do not correspond to the Sauter field.

The axion domain wall profile the reference \cite{paper4} employs is given by
\begin{equation}
    a(r)=  \frac{A R_e}{r} \tanh\gamma (r-R_0),
    \label{spha}
\end{equation}
 \begin{equation}
     a(r)= A \tanh\gamma (r-R_0+\delta).
     \label{spha2}
 \end{equation}
The formula  (\ref{spha}) corresponds to the regime $R_{trans}>r>R_0$ and (\ref{spha2}) to the regime $r>R_{trans}$, where the constants $\gamma=1/2 m_a$ y $A=\pi/2 f_a$ have been introduced.
The value $R_{trans}$ is given in \cite{paper4}.
For simplicity, assume that the magnetic field inside a nugget has a constant value and direction. In spherical coordinates, such magnetic field is given by
 \begin{equation}
     B= B( \cos \theta \;\hat{r} - \sin \theta\; \hat{\theta}).
     \label{b}
 \end{equation}
 Given a magnetic field and an axion wall, an electric field $E$ is induced.  The Poisson equation for the potential $V$ such that $\nabla V=-E$ that arises from the term $\nabla (aB)$ in (\ref{1}) is expressed in the form
 \begin{equation}
     \small{\Delta V= g A B R_e \Bigg[\frac{\cos \theta }{r^2} \Bigg( \tanh\gamma(r-R_0)+\gamma r \text{sech}^2\gamma(R_0-x) \Bigg)-   \frac{\cot \theta}{r^2} \tanh\gamma (r-R_0) \Bigg],
     \label{poi1}}
 \end{equation} 
 \begin{equation}
     \small{\Delta V=g A B \Bigg[ \frac{\cos \theta}{r^2} \bigg(2 r \tanh\gamma(r-R_0+\delta)+\gamma r^2 \text{sech}^2\gamma(R_0-\delta-x) \bigg) -  \frac{\cot \theta}{r}    \tanh\gamma (r-R_0+\delta) \Bigg],
     \label{poi2}}
 \end{equation}
 the first expression corresponds to  $R_{trans}>r>R_0$ and the second $r>R_{trans}$.
 We can solve both equations using the Green function method, but due to the fact that the effective charge is complicated  multipole expansion is required. Due to the azimuthal symmetry, the Green function becomes 
 \begin{equation}
     G(x,x')= 4 \pi \sum_{l=0}^{\infty} \frac{1}{2l+1} \frac{r^l_{<}}{r'^{l+1}_{>}} Y_{l,0}^{+}(\theta') Y_{l,0}(\theta),
     \label{mult}
 \end{equation}
 where $r_>$ and $r_<$ denotes the biggest and smallest respectively between $r$ and $r'$. One of the boundary conditions is that the potential goes to zero at infinity.

 The monopole therm is zero for both equations. In this section, the dipole case is worked out explicitly, the correction terms for the general case are given in the next one.  
In the region $R_0<r<R_{trans}$  the solution is of the form 
  \begin{equation}
   V=g A B \pi \cos \theta \bigg[\frac{Q_1}{r^2}+r (Q_2+Q_3)\bigg],
    \label{solucio1}
 \end{equation}
where $Q_1$, $Q_2$ y $Q_3$ denotes integrals that come from the moments of the field  given by
\begin{equation}
    Q_1=R_e\int_{R_0}^r \bigg[\bigg(\frac{2}{3}-\frac{\pi}{2}\bigg) x \tanh\gamma(x-R_0)+\gamma x^2 \text{sech}^2\gamma(R_0-x) \bigg]dx,
    \label{int1}
\end{equation}
\begin{equation}
    Q_2=R_e \int_{r}^{R_{trans}} \bigg[\bigg(\frac{2}{3}-\frac{\pi}{2}\bigg) \frac{\tanh\gamma(x-R_0)}{x^2}+ \frac{\gamma\text{sech}^2\gamma(R_0-x)}{x} \bigg]dx,
    \label{int2} 
\end{equation}
\begin{equation}
    Q_3=\int_{R_{trans}}^{R} \bigg[\bigg(\frac{4}{3}-\frac{\pi}{2}\bigg) \frac{\tanh\gamma(x-R_0+\delta)}{x} + \frac{2 \gamma}{3} \text{sech}^2\gamma(x-R_0+\delta)\bigg] dx.
    \label{int3}
\end{equation}
In the region $R_{trans}<r<R$ the resulting potential is 
\begin{equation}
    V=g A B \pi \cos \theta \bigg(\frac{Q_4+Q_5}{r^2}+r  Q_6 \bigg),
    \label{sol2}
\end{equation}
In this case the integrals $Q_4$, $Q_5$ y $Q_6$ are described by
\begin{equation}
    Q_4=R_e \int_{R_0}^{R_{trans}} \bigg[\bigg(\frac{2}{3}-\frac{\pi}{2}\bigg) x \tanh\gamma(x-R_0)+ \frac{2 \gamma}{3} x^2 \text{sech}^2\gamma(x-R_0)\bigg] dx,
    \label{int4}
\end{equation}
\begin{equation}
    Q_5=\int_{R_{trans}}^{r} \bigg[\bigg(\frac{4}{3}-\frac{\pi}{2}\bigg) x^2 \tanh \gamma(x-R_0+\delta)+ \frac{2\gamma}{3} x^3 \text{sech}^2\gamma(x-R_0+\delta)\bigg] dx,
    \label{int5}
\end{equation}
\begin{equation}
    Q_6=\int_{r}^{R}  \bigg[\bigg(\frac{4}{3}-\frac{\pi}{2}\bigg)\frac{\tanh\gamma(x-R_0+\delta)}{x} + \frac{2 \gamma}{3} \text{sech}^2\gamma(x-R_0+\delta) \bigg]dx.
    \label{int6}
\end{equation}
The integrals described in each solution can't be solved by analytic means in general. Therefore, some reasonable approximations are needed for solving them.
For the first integral (\ref{int1}), one may use the expansion $$\tanh(x-b)\sim 1-2 e^{-2(x-b)},$$ and that, for $\gamma(R_{trans}-R_0)<<1$
$$\text{sech}^2a(x-b)=1-a^2(x-b).$$ By use of this, it is obtained that
\begin{equation}
    Q_1=\frac{R_e}{\gamma^2}\Bigg[\frac{2 }{3}-\frac{\pi}{2}\Bigg]\Bigg[e^{2(R_0 \gamma-x\gamma)}\big(x\gamma+1/2\big)+\frac{ (x\gamma)^2}{2}\Bigg]\Big |_{R_0}^r
    +\frac{R_e}{18} \Bigg[(\gamma x)^3\Big(4 \gamma-3\gamma x+4\Big)\Bigg] \Big|_{R_0}^{r}.
    \label{q1}
\end{equation}
For the second equation we are gonna use the Laurent series of the hyperbolic tangent and truncate the series to get  $$\frac{\tanh(x-b)}{x^2}= \frac{-\tanh(b)}{x^2}+\frac{\text{sech}^2b}{x},$$ and, like the first equation, we are gonna use $\text{sech}^2(x-b)\sim \text{sech}^2 b$, both approximations apply in the regime $a(R_{trans}-R_0)<<1$. With these approximations, it is found 
\begin{equation}\label{q2}
   \begin{gathered}
        Q_2 = R_e a \big[\frac{2}{3}-\frac{\pi}{2}\big] \Bigg[\tanh(\gamma R_0) \frac{1}{\gamma x} +  \text{sech}^2(R_0 \gamma) \log(\gamma x)\Bigg]\Bigg|_r^{R_{trans}} + \Bigg[\frac{2 \gamma}{3}R_e\log(x)\text{sech}^2(b\gamma) \Bigg]\Bigg|_r^{R_{trans}}.
   \end{gathered}
\end{equation}
For the third integral, one may take into account that  $\tanh(x-b)/x\simeq 1/x$, for a sufficiently large nugget radius, $\gamma R>1$. A direct integration leads to
\begin{equation}
    Q_3=\Big[\frac{4}{3}-\frac{\pi}{2}\Big] \Big[\log(\gamma x)\Big]\Big|_R^{R_{trans}} +(2/3)\Big[\tanh\gamma (x-R_0+\delta)\Big]\Big|_{R_{trans}}^R.
    \label{q3}
\end{equation}
The next integral, can be solved using the same tricks we used for the first one, the result is
\begin{equation}\label{q4}
   \begin{gathered}
     Q_4=\frac{R_e}{\gamma^2}\Bigg[\frac{2 }{3}-\frac{\pi}{2}\Bigg]\Bigg[e^{2\gamma (R_0 -x )}\big(x \gamma+1/2\big)+\frac{ (x \gamma)^2}{2}\Bigg]\Big|_{R_0}^{R_{Trans}}
    -\frac{R_e}{18 a^2} \Bigg[(\gamma x)^3\Big((3\gamma x-4 \gamma R_0-4\Big)\Bigg] \Big|_{R_0}^{R_{Trans}}.
   \end{gathered}
\end{equation}
The fifth one may be found by the approximation $\tanh(x-b)\sim 1-2 e^{-2(x-b)}$, and by the expansion $\text{sech}^2(x-b)=4 e^{2(b-x)}(1-e^{2(b-x)})$. Both approximations lead to the
expression
$$
 Q_5= \frac{1}{6 \gamma^3}\bigg(\frac{4}{3}-\frac{\pi}{2}\bigg) \Big[ e^{2(\gamma\big(R_0+\delta-x)\big) \big(6(\gamma x)^2+6\gamma x+3\big)+2x^3}\Big]\bigg|_{R_{trans}}^r
$$
\begin{equation}\label{q5}  
    +  \frac{e^{2\gamma (R_0+\delta - 2x)}}{24 \gamma^3} 
 \Bigg[e^{2\gamma (R_0+\delta)} \bigg(32 (\gamma x)^3+24 (\gamma x)^2 + 12 (\gamma x) + 3\bigg)+ 8 e^{2\gamma x} \bigg(4 (\gamma x)^3+6 (\gamma x)^2 +6 (\gamma x)+3 \Bigg] \bigg|_{R_{trans}}^r 
\ee
The last one can be integrated using the same trick as the ones used in the third one. The result of this procedure yields
\begin{equation}\label{q6}
    Q_6=\bigg(\frac{4}{3}-\frac{\pi}{2}\bigg) \Big[\log(\gamma x)\Big]\bigg|_r^{R_{trans}} +\frac{2}{3}\tanh\big(\gamma (x-R_0+\delta)\big) \bigg|_{r_{trans}}^R.
\end{equation}

\subsection{Higher order corrections}

Having the functional form of the potential for the dipolar case is not the end of the story, due to the fact that there are gonna be higher order correction due to the fact that $\cot\theta$ can be expanded in an infinite series of Legendre functions.
The solution of the potential in the regime $r>R_{trans}$ at all orders is given by
$$
 V(r,\theta)= 8 \pi^2 A B g \sum_{l=0}^\infty \frac{Y_l^0(\theta)}{2l+1} \Bigg\{  \int_0^\pi  Y_l^0 \sin \theta\cos\theta d\theta \Bigg[ \frac{1}{r^{l+1}}\int_{R_{trans}}^r \big[ 2 x^{l+1} \tanh\gamma(x-R_0+\delta)
  $$
 $$
  +\gamma x^{l+2} \text{sech}^2\gamma(R_0-\delta-x) \big] dx  
  +\frac{1}{r^{l+1}} \int_{R_0}^{R_{trans}} R_{eff} x^{l+2} \bigg[ \frac{ \tanh\gamma(x-R_0)}{x^2} + \frac{\gamma \text{sech}^2(R_0-x)}{x} \bigg] dx 
  $$
  $$
   + r^l  \int_r^R [2 x^{-l} \tanh\gamma(x-R_0+\delta) +\gamma x^{-l} \text{sech}^2\gamma(R_0-\delta-x)] dx\bigg]
  $$
 $$
 - \int_0^\pi Y_l^0 \sin\theta \cos\theta d\theta \Bigg[ \frac{1}{r^{l+1}} 
  \int_{R_0}^{R_{trans}} R_{eff} x^l \tanh\gamma(x-R_0) dx 
  $$
  \begin{equation}
  +\frac{1}{r^{l+1}}  \int_{R_{trans}}^r x^{l+1}\tanh\gamma(x-R_0) dx
  + r^l \int_r^R x^{-l} \tanh\gamma(x-R_0+\delta) dx \Bigg] \Bigg\}.
    \label{mlp}
\end{equation}
This expansion includes the terms for the dipolar case found in the previous section, plus higher orders. First, let notice that the dipolar order is important due to the fact that $\cos\theta$ is proportional to  $Y_1^0$ and will be orthogonal to the rest of the spherical harmonics. Second, let's also notice that $\cot \theta$  goes to $\infty$ at 0 and to $-\infty$ at $\pi$, thus infinite terms in the expansion are needed.

The higher order corrections, when the approximation $\tanh(x-b)=1-2 e^{-2(x-b)}$ was taken into account, are given by
$$
    \Delta V(r,\theta) = -8 \pi^2 A B g \sum_{l=2}^\infty \frac{Y_l^0(\theta)}{2l+1} \int_0^\pi \bigg(Y_l^0 \tanh\theta' \cos \theta'\bigg) d\theta' \Bigg\{  \frac{1}{\gamma^{l+2} r^{l+1}} \Bigg[e^{2\gamma(R_0-\delta)} 2^{-l-1} \Gamma(l+2,2x)+ \frac{x^{l+2}}{l+2} \Bigg] \Bigg|_{\gamma r}^{\gamma R}
    $$ 
 \begin{equation}   
    +\gamma^{l-1} r^l \Bigg[ e^{2 \gamma (R_0-\delta)} 2^l \Gamma(1-l,2x) - \frac{x^{1-l}}{l-1}\Bigg]\Bigg|_{r\gamma}^{R\gamma} \Bigg\}.
    \label{correc}
\end{equation}
Here $\Gamma(a,b)$ denotes the incomplete gamma function.
The corrections for the other region $R_0<r<R_{trans}$ are similar to the ones here and we will not discuss here the functional form here.

\subsection{Schwinger effect for the non ferromagnetic axion quark nugget}

After finding approximate expression for the electric field induced due to the axion nugget, the next step is to discuss the decay due to Schwinger effect. The following discussion relies particularly on the work \cite{kleinert2}, and it will be applied to a  dipole type of electric field outside the nugget and a constant field inside. 
For simplicity, scalar field production will be considered. The Klein-Gordon equation for a pure electric field, by taking into account that $A_0=V$ and $A_i=0$
reads as follows
\begin{equation}
  [\partial_\mu \partial^\mu + m^2- 2 i q A_0 \partial_0 - q^2 A_0 A_0] \Phi =0.
    \label{kgm}
\end{equation}
In the following, a reduced one dimensional problem in one direction will be considered. This reduction is valid if the three dimensional electric fields $E(x,y,z)$ and the potential $V(x,y,z)$ are such that the tunneling length $a=z_+-z_-$ to be defined below is smaller
than the variation $\delta x_\bot$ of the potential $V$ the $xy$ plane. In other words, $a^{-1}>V^{-1}\partial_{x_\bot}V$.
%

Taking only the radial tunneling has two implications. First the angular part of the potential is discarded and only a radial part $V(r)$ is taken into account. In this situation, the ansatz employed in the literature \cite{kleinert2} is the following one
\begin{equation}
    \Phi= e^{i \epsilon t} \frac{\psi_l(r)}{r} \Omega(\theta) e^{im\varphi}.
    \label{armon}
\end{equation}
In the previous section, some approximated expressions for the electric potential of the nugget were found. It is difficult to obtain results
with these expressions. For simplickty
the potential to be employed is a dipole type one which, in one dimension, reads as follows
\begin{equation}
    V(z)=
    \left\{ \begin{array}{ll}
        V_0 z & z<R \\
        V_0 \frac{R^3}{z^2} & z>R
    \end{array} \right.
    \label{v}
\end{equation}
By inserting (\ref{armon}) into (\ref{armon}), it is obtained that
\begin{equation}
    [\partial^2_z + P_\bot^2  + m^2 c^2 -\frac{1}{c^2} [\epsilon- qV(z)]^2] \phi_{p_\bot,\epsilon}(z)=0.
    \label{sch2}
\end{equation}
This equation in one dimension can be solved by the WKB method. 
The transmission probability is
\begin{equation}
    W_{wkb}(p_\bot,\epsilon)=e^{\frac{-2}{\hbar}\int_{z-}^{z+} \sqrt{p_\bot^2 + m^2 c^2 - \frac{1}{c}[\epsilon-q V(z)]^2} dz}.
    \label{trans}
\end{equation}
Notice that the exponential factor is the momentum along the z direction, and this quantity goes to zero at the turning points.
This equation can be expressed as follows
\begin{equation}
    W_{wkb}(p_\bot,\epsilon)=e^{\frac{-2 E_c}{ E_0}[1+\frac{(c p_\bot)^2}{m^2 c^4}] \Big[ \frac{E_0}{\hbar E_c} \frac{m_e^2 c^4}{m_e^2 c^4 + p_\bot^2 c^2} \int_{z-}^{z+} \sqrt{p_\bot^2 + m^2 c^2 - \frac{1}{c}[\epsilon-q V(z)]^2} dz \Big]} 
    \label{dim}
\end{equation}
It is convenient to denote the exponent of the last expression as 
\begin{equation}
    G(p_\bot, \epsilon)= \frac{E_0}{\hbar E_c} \frac{m_e^2 c^4}{m_e^2 c^4 + p_\bot^2 c^2} \int_{z-}^{z+} \sqrt{p_\bot^2 + m^2 c^2 - \frac{1}{c^2}[\epsilon-q V(z)]^2} dz
    \label{g}
\end{equation}
In the general case, it is difficult to make to calculate $G$ explicitly. However
from (\ref{trans}) and (\ref{dim}) it is seen that  the critical field $E_c=\frac{m_e^2 c^3}{e \hbar}$ is proportional to $\hbar^{-1}$, so the most important contribution of $p_\bot$ is when $p_\bot ~ \sqrt{\hbar}$. 
Then the integral can be approximated as
$$
G(p_\bot. \epsilon)=G(0, \epsilon)+G_\delta(0, \epsilon)\delta+..
$$
with the dimensionless variable  $\delta= \frac{(c p_\bot)^2}{m_e^2 c^4}$.  In the present case, as the potential is split, 
the function $G$ is decomposed as $G=G_1+G_2$ with
\begin{equation}
    G_1(p_\bot, \epsilon)= \frac{E_0}{\hbar E_c} \frac{m_e^2 c^4}{m_e^2 c^4 + p_\bot^2 c^2} \int_{z-}^{R} \sqrt{p_\bot^2 + m^2 c^2 - \frac{1}{c^2}[\epsilon-q  V_0 z]^2} dz,
    \label{g1}
\end{equation}
\begin{equation}
    G_2(p_\bot, \epsilon)= \frac{E_0}{\hbar E_c} \frac{m_e^2 c^4}{m_e^2 c^4 + p_\bot^2 c^2} \int_{R}^{z+} \sqrt{p_\bot^2 + m^2 c^2 - \frac{1}{c^2}[\epsilon-q V_0 \frac{R^3}{z^2}]^2} dz.
    \label{g2}
\end{equation}
In the last expressions, the points $z_+$ and $z_-$ denote the turning points, defined by the condition $E_+(p_z=0, p_\bot, z_+)=E_-(p_z=0, p_\bot, z_-)=\epsilon$. Both point share the same energy. 
Equivalently, they are given by 
\begin{equation}
    V(z_\pm)=\mp \sqrt{c^2 p_\bot^2+m^2 c^4} + \epsilon.
    \label{turn}
\end{equation}
The only problem we can face with this is the fact that $z_+$ is squared in the potential, but we can use the fact that $z_+>R>0$ 
in order to obtain
\begin{equation}
    z_-=\frac{ \sqrt{c^2 p_\bot^2+m_e^2 c^4}+\epsilon}{q V_0},
    \label{z-}
\end{equation}
\begin{equation}
    z_+=\sqrt{\frac{V_0 R^3 q}{\epsilon - \sqrt{p^2 c^2+m_e^2 c^4}}}.
    \label{z+}
\end{equation}
The expressions (\ref{g1}) and (\ref{g2}) can be integrated to the lowest order $\delta=0$, the result is 
\begin{equation}
    G_1= \frac{E_0}{\hbar E_c} \frac{m_e^2 c^4}{m_e^2 c^4 + p_\bot^2 c^2} \frac{q v_0}{c} \Bigg[ \frac{1}{8} \big(4c+d^2\big) \tan^{-1} \big( \frac{d-2z}{2 \sqrt{c+z(d-z)}} \big) \big)-\frac{1}{4} \big(d-2z\big) \sqrt{c+z\big(d-z\big)} \Bigg] \Bigg|_{z_-}^R.
    \label{r1}
\end{equation}
In (\ref{r1}) the variables $c= \frac{p_\bot^2 c^2+m_e^2 c^4 - \epsilon^2}{(q v_0)^2}$ and $d= \frac{2 \epsilon}{q v_0}$ were introduced.
The second contribution is
$$
 G_2= -\frac{E_0}{\hbar E_c} \frac{m_e^2 c^4}{m_e^2 c^4 + p_\bot^2 c^2} \frac{q v_0 R^3}{c}\Bigg\{ z \sqrt{a+\frac{b z^2-1}{z^4}} \Big[i \sqrt{2} z \sqrt{4 a+ b^2} \sqrt{\frac{-\sqrt{4a+b^2}+2 a z^2 + b}{b-\sqrt{4a+b^2}}} 
$$
$$    
\times \sqrt{\frac{\sqrt{4a+b^2}+2 a z^2 + b}{b+\sqrt{4a+b^2}}} F\bigg[ i \sinh^{-1}\bigg(\sqrt{2} \sqrt{\frac{a}{b+\sqrt{b^2+4a}}z}\bigg) \Big| \frac{b+\sqrt{b^2+4a}}{b-\sqrt{b^2+4a}} \bigg] - iz \big(\sqrt{4a+b^2}-b\big) 
$$
$$
\times \sqrt{\frac{-2\sqrt{4a+b^2}+4 a z^2 + 2b}{b-\sqrt{4a+b^2}}} \sqrt{\frac{\sqrt{4a+b^2}+2 a z^2 + b}{b+\sqrt{4a+b^2}}}
$$
  $$
    \times E\bigg[ i \sinh^{-1}\bigg(\sqrt{2} \sqrt{\frac{a}{b+\sqrt{b^2+4a}}z}\bigg) \Big| \frac{b+\sqrt{b^2+4a}}{b-\sqrt{b^2+4a}} \bigg] + 2 \sqrt{\frac{a}{\sqrt{4a+b^2}+b}} \big(a z^4+b z^2-1\big) \Big] \bigg\} 
    $$
\begin{equation}    
    \times \Bigg\{ \frac{1}{2 \sqrt{\frac{a}{\sqrt{4a+b^2}+b}} \big(a z^4+b z^2-1\big)} \Bigg\} \Bigg|_{R}^{z_+}
 \label{r2}
\end{equation}
Here $a= \frac{p_bot^2 c^2+m_e^2 c^4 - \epsilon^2}{(q v_0 R^3)^2}$ and $b= \frac{2 \epsilon}{q v_0 R^3}$. In these terms the event density in four space is given by
\begin{equation}
    \frac{d^4 N_{wkb}}{dt dx dy dz} =D_s  \frac{q^2 E_0 E(z)}{8 \pi^3 G_0\big(0,E(z)\big)} e^{-\pi(E_c/E_0)G(0,E(z)} ,
    \label{No}
\end{equation}
with $D_s$ the spin degeneration.

The formula (\ref{No}) takes into account only an electric field. But these methods can be generalized when a magnetic field is present, if one assumes for simplicity that the magnetic field is constant \cite{kleinert2}. The magnetic field in a magnetic nugget is gonna be parallel to the electric field.
The resulting event density is given by
\begin{equation}
    \frac{d^4 N_{wkb}}{dt dx dy dz} =D_s f_{0,1/2}\bigg(\frac{B G_0(0,E(z))}{E_0}\bigg)  \frac{q^2 E_0 E(z)}{8 \pi^2 G_0\big(0, E(z)\big)} e^{-\pi(E_c/E_0)G(0,E(z))} .
    \label{N}
\end{equation}
Here the function
\begin{equation}
     f(x)=
    \left\{ \begin{array}{ll}
        \frac{\pi x}{\sinh(\pi x)} & \sigma=0 \\
        2 \cosh(\frac{\pi g x}{2}) \frac{\pi x}{\sinh(\pi x)} & \sigma=1/2
    \end{array} \right.
    \label{part}
\end{equation}
has been introduced. The limit $B\to0$ the formula (\ref{No}) is recovered.


\begin{thebibliography}{99}
\bibitem{zhitnitsky} A. Zhitnitsky JCAP 0310 (2003) 010.
\bibitem{zhitnitsky1} D. Oaknin and A. Zhitnitsky Phys.Rev. D71 (2005) 023519.
\bibitem{zhitnitsky2} A. Zhitnitsky Phys.Rev. D74 (2006) 043515.
\bibitem{zhitnitsky3} K. Lawson and A. Zhitnitsky Phys. Lett. B 724 (2013) 17.
\bibitem{zhitnitsky4} K. Lawson and A. Zhitnitsky Phys. Rev. D 95 (2017) 063521.
\bibitem{zhitnitsky5} X. Liang and A. Zhitnitsky Phys. Rev. D 94 (2016) 083502.
\bibitem{zhitnitsky6} S. Ge, X. Liang and A. Zhitnitsky Phys. Rev. D 97 (2018) 043008.
\bibitem{zhitnitsky7} A. Zhitnitsky Physics of the Dark Universe 22 (2018), 1
\bibitem{zhitnitsky8}  K. Lawson and A. Zhitnitsky Physics of the Dark Universe (2019) 100295.
\bibitem{zhitnitsky9} N. Raza, L. van Waerbeke and A. Zhitnitsky Phys. Rev. D 98 (2018) 103527.
\bibitem{zhitnitsky10} H. Fischer, X.Liang, Y. Semertzidis, A. Zhitnitsky and K. Zioutas Phys. Rev. D 98 (2018) 043013.
\bibitem{zhitnitsky11} L. van Waerbeke and A. Zhitnitsky Phys. Rev. D 99 (2019) 043535.
\bibitem{zhitnitsky12} V. Flambaum and A. Zhitnitsky Phys. Rev. D 99 (2019) 043535.
\bibitem{zhitnitsky122} K. Lawson and A. Zhitnitsky JCAP 02  (2017) 049.
\bibitem{zhitnitsky13} S. Ge, K. Lawson and A. Zhitnitsky Phys. Rev. D 99 (2019) 116017.
\bibitem{tatsumi} T. Tatsumi Phys. Lett. B 489 (2000) 280.
\bibitem{son} D. Son and A. Zhitnitsky Phys. Rev. D70 (2004) 074018.
\bibitem{son1}  D. Son and M. Stephanov Phys. Rev. D 77 (2008) 014021.
\bibitem{son2} M. Melitski and A. Zhitnitsky Phys.Rev. D72 (2005) 045011.
\bibitem{nature}  J. Pace VanDevender, A. VanDevender, T. Sloan, C. Swaim, P. Wilson, R.Schmitt, R. Zakirov, J. Blum, J. Cross Sr. and Niall McGinley Scientific Reports 7 (2017) 8758.
\bibitem{bodmer} A. R. Bodmer, Phys. Rev. D4  (1971) 1601.
\bibitem{witten} E. Witten Phys. Rev. D 30 (1984) 272.
\bibitem{atreya} A. Atreya, A. Sarkar. and A. Srivastava Phys. Rev. D 90 (2014) 045010.
\bibitem{bjorken1} S. A. Chin and A. K. Kerman, Phys. Rev. Lett. 43 (1979) 1292.
\bibitem{bjorken2} J. D. Bjorken and L. D. McLerran, Phys. Rev. D20 (1979) 2353.
\bibitem{estable1} C. Alcock and E. Farhi, Phys. Rev. D 32 (1985) 1273.
\bibitem{estable2} A. Bhattacharyya  et al. Phys. Rev. D 61 (2000) 083509.
\bibitem{estable3} G. Lugones and J. Horvath Phys. Rev. D 69 (2004) 063509.

\bibitem{madsen} J. Madsen Phys. Rev. Lett 61 (1988) 2909.

\bibitem{madsen2} J. Madsen Nucl. Phys. B 24 (1991) 84.

\bibitem{estable4} J. Madsen and H. Heiselberg and K. Riisager, Phys. Rev. D 34 (1986) 2947.
\bibitem{estable5} C. Alcock and A. Olinto, Phys. Rev. D 39 (1989) 1233.
\bibitem{estable6} K. Sumiyoshi and T. Kajino , in Proceedings of the International Workshop on Strange Quark Matter in Physics
and Astrophysics, eds. J. Madsen and P. Haensel, Nucl.
Phys. B. Proc. Supp.24, 80 (1991).
\bibitem{estable8} J. Madsen and M. Olesen, Phys. Rev. D 43 (1991) 1069.
\bibitem{estable9} M. Olesen and J. Madsen, Phys. Rev. D 47 (1993) 2313.
\bibitem{estable10} L. Masperi and M. Orsaria, Physics of Particles and Nuclei Letter 1 (2004) 48.
\bibitem{applegate} J. Applegate and C. Hogan Phys. Rev. D 31 (1985) 3037.
\bibitem{schaeffer} R. Schaeffer, P. Delbourgo-Salvadora and J. Audouze Nature 317 (1985) 407. 

\bibitem{bloch} F. Bloch Z. Phys. 57 (1929) 545.
\bibitem{lam} J. Lam Phys. Kondens. Materie 15 (1972) 46.
\bibitem{chakrabarty}  S. Chakrabarty Phys. Rev. D 54 (1996) 1306.

\bibitem{sikivie} P. Sikivie Lect. Notes Phys. 741 (2008) 19.
\bibitem{preskill} J. Preskill, M. Wise and F. Wilczek, Phys. Lett. B 120 (1983) 127.
\bibitem{preskilo} M. Dine and W. Fischler Phys.Lett. B120 (1983) 137.
\bibitem{preskill2} L.Abbot and P. Sikivie, Phys. Lett. B120 (1983) 133.
\bibitem{dicus}  D. Dicus, E. Kolb, V. Teplitz, and R. Wagoner Phys. Rev. D18 (1978) 1829. 
\bibitem{bardeen} W. Bardeen and S. Tye, Phys. Lett. 74B (1978) 580.
\bibitem{peccei} R. Peccei and H. Quinn, Phys. Rev. Lett. 38 (1977) 1440; Phys. Rev. D 16 (1977) 1791.
\bibitem{dfsz}  M. Dine, W. Fischler and M. Srednicki Phys. Lett. B 104 (1981) 199; A. Zhitnitsky Sov. J. Nucl.
Phys. 31 (1980) 260.
\bibitem{svz} M. Shifman, A. Vainstein and V. Zakharov, Nucl. Phys. B 166 (1980) 493; J. Kim Phys. Rev.
Lett. 43 (1979) 103.
\bibitem{lyth}  D. Lyth and E. Stewart, Phys. Rev. D 46 (1992) 532; Phys. Lett. B283 (1992) 189.
\bibitem{lyth2} Planck Collaboration: P. A. R. Ade et all, Astron. Astrophys. 571 (2014) A22.
\bibitem{lesgourgues} M.Beltran, J. Garcia-Bellido and J. Lesgourgues Phys. Rev. D 75 (2007) 103507.
\bibitem{cfl}  M. Alford, K. Rajagopal, T. Schäfer and A. Schmitt Reviews of Modern Physics. 80 (2008) 4.
\bibitem{cfl2} M. Alford, K. Rajagopal and F. Wilczek Physics Letters B. 422 (1998) 247.



\bibitem{dunne1} G. V. Dunne, C. Schubert  Phys. Rev. D 72 (2005) 105004.
\bibitem{dunne2} G. V. Dunne, H. Gies, C. Schubert and Q. Wang  Phys. Rev. D 73 (2006) 065028.
 \bibitem{dunne3} G. V. Dunne, Q. Wang Phys. Rev. D 74 (2006) 065015.
\bibitem{dunne4} D. D. Dietrich, G. V. Dunne J. Phys. A 40 (2007) F825.
\bibitem{dunne5} C. K. Dumlu, G. V. Dunne Phys. Rev. D 84 (2011) 125023.

\bibitem{ilderton}  A. Ilderton, G. Torgrimsson and J. Wardh Phys. Rev. D 92 (2015) 065001.
\bibitem{ilderton2} A. Ilderton, G. Torgrimsson and J. Wardh Phys. Rev. D 92 (2015) 025009.
\bibitem{ilderton3} A. Ilderton  JHEP 09 (2014) 166.
\bibitem{ilderton4} F. Hebenstreit, A. Ilderton and M. Marklund Phys. Rev. D. 84 (2011) 125022.

\bibitem{kim} S. Kim and D. Page  Phys.Rev. D65 (2002) 105002.
\bibitem{kim2} S. Kim and D. Page Phys. Rev. D 73 (2006) 065020.
\bibitem{nikishov} A. I. Nikishov, Sov. Phys. JETP 30 (1970) 660 (1970); Nucl. Phys. B 21 (1970) 346.
\bibitem{kleinert1} A. Chervyakov and H. Kleinert Phys. Rev. D 80 (2009) 065010.
\bibitem{kleinert2} H. Kleinert, R. Ruffini and S. Xue Phys .Rev. D 78 (2008) 025011.
\bibitem{tanji} M. Tanji Ann. Phys. 324 (2009) 1691.
\bibitem{su} W.  Su, M. Jian, Z. Lv, R. Grobe and Q. Su Phys. Rev. A 86 (2012) 013422.

\bibitem{expansion1} R. Ruffini, G. Vereshchagin, and S. Xue, Phys. Rep. 487 (2010) 1.
\bibitem{expansion2} R. Ruffini and S. Xue, J. Korean Phys. Soc. 49 (2006) S715.
\bibitem{expansion3} H. Kleinert, E. Strobel, and S.-S. Xue, Phys. Rev. D. 88 (2013) 025049.
\bibitem{karbstein1} F. Karbstein Phys. Rev. D 88 (2013) 085033.



\bibitem{karbstein33}  F. Karbstein, L. Roessler, B. Döbrich and H. Gies Int. J. Mod. Phys. Conf. Ser. 14 (2012) 403.

\bibitem{karbstein7}  F. Karbstein and R. Shaisultanov Phys. Rev. D 91 (2015) 113002.

\bibitem{karbstein11} T. Adorno, A. Gitman and D. Shabad "Magnetic response from constant backgrounds to Coulomb sources" arXiv:1710.00138.


\bibitem{karbstein2} B.Döbrich, H. Gies, N. Neitz and F. Karbstein Phys. Rev. D 87 (2013) 025022.
\bibitem{karbstein3} B.Döbrich, H. Gies, N. Neitz and F. Karbstein Phys. Rev. Lett. 109 (2012) 131802.
\bibitem{karbstein6} H. Gies, F. Karbstein and C. Kohlfürst Phys. Rev. D 97 (2018) 036022.

\bibitem{karbstein10} J. Koga, M. Murakami, A. Arefiev and Y. Nakamiya Matter and Radiation at Extremes 4 (2019) 034401. 

\bibitem{karbstein8}  S. Evans and J. Rafelski Phys. Lett. B 791 (2019) 331.
\bibitem{karbstein5} F. Karbstein "The quantum vacuum in electromagnetic fields: From the Heisenberg-Euler effective action to vacuum birefringence " in "Quantum Field Theory at the Limits: from Strong Fields to Heavy Quarks" July 2016.
\bibitem{karbstein4}  H. Gies and F. Karbstein JHEP 03 (2017) 108.


\bibitem{karbstein9} P. Ripoche and J. Heyl  Phys. Rev. D 99 (2019) 083004.


\bibitem{rujula} A. De Rujula and S. Glashow Nature 312 (1984) 734.
\bibitem{burdin} S. Burdin, M. Fairbairn, P. Mermod, D. Milstead, J. Pinfold, T. Sloan, W. Taylor Phys. Rep. 582 (2015) 1.
\bibitem{porter} N. Porter, M. Cawley, D. Fegan, G. MacNeill and T. Weekes Irish Astron. J. 18, 193–196 (1988).
\bibitem{auger} Abraham, J. et al. Nucl. Instr. and Meth. Phys. Res. A620 (2010) 227.
\bibitem{icecube} IceCube Collaboration M. Aartsen et al. Eur. Phys. J. C 75
(2015) 492.
\bibitem{cth} M. McGlaun, S. Thompson and M. Elrick, CTH: A three-dimensional shock wave physics code. Int. J. Impact Eng. 10  (1990) 351.
\bibitem{sesame} S. Crocket  SESAME Database: Equation-of-State tabular data for the thermodynamic properties of materials. http://www.lanl.gov/
org/padste/adtsc/theoretical/physics-chemistry-materials/sesame-database.php, (1999) (Date of access: 24/06/2017).
\bibitem{alphamagnetic1} J. Madsen Phys. Rev. D 71 (2005) 014026.
\bibitem{alphamagnetic2}  J. Madsen J. Phys. G 31 (2005) S833.
\bibitem{alphamagnetic3} M. Aguilar et al. [AMS Collaboration], Phys. Rept. 366  (2002) 331.
\bibitem{alphamagnetic4}  J. Sandweiss, J. Phys. G 30, S51 (2004).
\bibitem{showers} K. Lawson Phys. Rev. D83 (2011) 103520.
\bibitem{telescopearray} H.Tokuno et al.,Nucl. Instrum. Meth. A 676 (2012) 54.
\bibitem{pshirkov} M. Pshirkov International Journal of Modern Physics D 25 (2016)1650103.
\bibitem{salamon} P. B. Price and M. H. Salamon, Phys. Rev. Lett. 56 (1986) 1226.
\bibitem{sasso} M. Ambrosio et al MACRO collaboration Eur. Phys. J. C 13 (2000) 453.
\bibitem{wandelt}  B. Wandelt. et al. Self-interacting dark matter. Ch. 5 in Sources and Detection of Dark Matter
and Dark Energy in the Universe, from 4th Intl. Symp., Marina del Rey, CA, USA, February
2325, 2000, (ed. Cline, D. B.) 263274. (Springer, 2001). https://arxiv.org/ abs/astro-ph/0006344,
(2000) (Date of access: 24/06/2017).
\bibitem{rafelski} J. Rafelski,  L. Labun and J. Birrell PRL 110 (2013) 111102.
\bibitem{paper4} X. Liang and A. Zhitnitsky Phys. Rev. D 99 (2019) 023015.
\bibitem{iwazaki} A. Iwazaki Phys. Lett. B 489 (2000) 353.

\end{thebibliography}
\end{document}